# Spectral Efficiency Optimization For Millimeter Wave Multi-User MIMO Systems

Qingjiang Shi and Mingyi Hong

*Abstract*—As a key enabling technology for 5G wireless, millimeter wave (mmWave) communication motivates the utilization of large-scale antenna arrays for achieving highly directional beamforming. However, the high cost and power consumption of RF chains stands in the way of adoption of the optimal fully-digital precoding in large-array systems. To reduce the number of RF chains while still maintaining the spatial multiplexing gain of large-array, hybrid precoding architecture has been proposed for mmWave systems and received considerable interest in both industry and academia. However, the optimal hybrid precoding design has not been fully understood, especially for the multi-user MIMO case. This paper is the first work that directly addresses the nonconvex hybrid precoding problem of mmWave multi-user MIMO systems (without any approximation) by using penalty dual decomposition (PDD) method. The proposed PDD method have a guaranteed convergence to KKT solutions of the hybrid precoding problem under a mild assumption. Simulation results show that, even when both the transmitter and the receivers are equipped with the fewest RF chains that are required to support multi-stream transmission, hybrid precoding can still approach the performance of fully-digital precoding in both the infinite resolution phase shifter case and the finite resolution phase shifter case with several bits quantization.

## I. Introduction

The frequency bandwidth scarcity has motivated the exploration of the underutilized millimeter wave (mmWave) frequency spectrum for future broadband cellular communication networks [1]–[5]. The shorter wavelength of the mmWave frequencies enables more antenna components to be packed in the same physical dimension, which allows for large-scale spatial multiplexing and highly directional beamforming. For massive multiple-input-multiple-output (MIMO) mmWave systems, the conventional fully-digital (FD) precoding scheme, which requires one radio frequency (RF) chain per antenna element, is not viable due to the high cost and the high power consumption of RF chain components in high frequencies. To address the challenge of this hardware limitation while exploiting the multiplexing gain of MIMO, hybrid precoding architectures have been proposed for mmWave systems and widely investigated in the literature [6]–[19].

This work was supported in part by NSFC under Grant 61671411, 61631020, 61374020, 61611130211, and by Zhejiang Provincial NSF of China under Grant LR15F010002. The work of M. Hong is supported in part by NSF under Grant CCF-1526078 and in part by AFOSR under Grant 15RT0767.

Q. Shi is with the College of EIE, Nanjing University of Aeronautics and Astronautics, Nanjing 210016, China. Email: qing.j.shi@gmail.com.
M. Hong is with the Dept. of ECE, University of Minnesota, Minneapolis, MN 55455, USA. Email: mhong@umn.edu.
This work was done when both authors worked at the Dept. of Industrial and Manufacturing Systems Engineering, Iowa State University, IA 50011, USA.

The key idea of hybrid precoding is using a linear network of variable phase shifters in the RF domain in addition to the baseband digital precoding. Such a scheme was firstly known as *soft antenna selection* (SAS) which was proposed in [20] to improve the performance of the traditional antenna selection scheme. It was shown in [20] that the SAS method can even achieve the optimal performance of a fully-digitally precoded single-stream MIMO system when each end is equipped with two or more RF chains. Recently, SAS is re-introduced to mmWave system design under the name of hybrid precoding/beamforming [2], [6], which has received significant interest in recent literature on large-scale antenna array systems.

The pioneering work on hybrid precoding for mmWave systems [6] investigated hybrid precoding methods for a point-to-point mmWave MIMO system. It was shown in [6] that the spectral efficiency maximization problem for mmWave MIMO systems can be approximately solved by addressing a matrix approximation or reconstruction problem, i.e., minimizing the Frobenius norm of the difference between the optimal FD precoder and the hybrid precoder. By exploiting the sparse nature of mmWave channels, the matrix approximation problem is reformulated as a compressive-sensing-like problem which is addressed by using a modified orthogonal matching pursuit (OMP) algorithm. The OMP method in [6] can achieve good performance when hundreds of antennas are used at both ends of transceiver and/or the number of RF chains is strictly greater than the number of data streams. But there is still a significant performance gap between the hybrid precoding method proposed in [6] and the optimal FD precoding method especially when the number of RF chains is equal to the number of data streams. Hence, several works have devoted to reducing this performance gap. The authors of [7] proposed using alternating optimization to approximate the solution of the matrix approximation problem. To deal with the difficulty arising from the unit-modulus constraints, [7] used manifold optimization to solve the analog (or RF) precoder design problem with fixed digital precoder. Other works that are based on matrix approximation can be found in [8]–[10]. Together with [6], [7], all the above methods don't directly take into consideration the power constraint in the optimization, they inevitably incur some performance loss as compared to the optimal hybrid precoding performance. Differently from the matrix approximation methods [6]–[10], the work [15] proposed another hybrid precoding algorithm that can approximately solve the spectral efficiency maximization problem of mmWave MIMO systems. First, by removing the coupling among the hybrid precoder and decoders (or combiners) using some approximations under large-antenna array, the analog precoder is designed independently. Then, given the analog precoder, the digital precoder, the analog combiner, and the digital combiner are sequentially designed. Simulations confirm that the hybrid precoding method in [15] can achieve a performance that is very close to the FD precoding performance, though it is still a heuristic algorithm.

While the above works considered hybrid precoding for single-user MIMO systems, there are a few works on hybrid precoding for spectral efficiency optimization of multi-user MIMO/MISO systems [11], [12], [15]–[18]. In [11], the authors developed a low-complexity two-stage hybrid precoding algorithm for downlink mmWave *single-stream* multi-user MIMO systems with limited feedback channel. In the first stage, the analog precoder at the base station (BS) and the analog combiners at the users are jointly designed to maximize the desired signal power of each user while neglecting the resulting interference among users. In the second stage, the BS digital precoder is designed to manage the multi-user interference. The authors of [12] proposed equal-gain-transmission-based analog beamforming combined with block diagonalization (BD) based digital precoding for generic channel setup. By exploiting the knowledge of angle of departure, the work [13] proposed iterative matrix decomposition based BD for mmWave multiuser MIMO systems. In [14], the authors first proposed a low complexity mmWave channel estimation algorithm for multiuser mmWave systems with single RF chain at each user and then designed a hybrid analog precoding and digital zero-forcing precoding scheme based on the channel estimations. In [15], assuming perfect channel state information, the authors proposed an iterative hybrid precoding algorithm for multi-user MISO systems, which iterates between the design of zero-forcing (ZF) precoder and the analog precoder. The works [16], [17] proposed low complexity hybrid precoding schemes for multi-user MISO systems by performing phase-only maximum ratio combining (MRC) or channel matched filter in the analog domain and ZF beamforming in the digital domain based on the effective channel. Extending the hybrid precoding methods in [16], [17], the work [18] proposed similar low complexity hybrid precoding algorithms for multi-user MIMO systems by designing the analog and digital precoders sequentially, where digital precoders are separated to pre-filters and post-filters. First, analog precoders and decoders are designed based on a per-user channel matching criterion. Then digital pre-filters are applied at both the transmitter side and the receiver side. Finally, optimal digital post-filters are derived based on four linear transmit strategies, including ZF [21], MMSE [22], block diagonalization (BD) [23], regularized BD (RBD) [24], followed by optimal power allocation via the water-filling algorithm. The simulation results in [18] show that the RBD method performs the best among various transmit strategies. Differently from the above works focusing on the downlink systems, the authors of [19] investigated the hybrid beamforming design problems for a mmWave massive multicell MIMO *uplink* transmission system, tackling both the intracell and inter-cell interference under the hardware constraints.

For tractability, most of the above works typically assume infinite resolution phase shifters used for analog precoders, although it is very expensive to realize accurate phase shifters [25], [26]. Usually, low resolution but less expensive phase shifters are used in practice, resulting in spectral efficiency optimization problems with *discrete* unit modulus constraints [6], [15], [17]. Since it is impossible to directly solve the discrete optimization problem, the most straightforward way to design analog precoder with finite resolution phase shifters is to address the spectral efficiency problem assuming infinite resolution first and then to quantize analog precoder using a finite set of available phases [17]. However, this approach could be ineffective for the case of very low resolution phase shifters. An alternative way to approximately addressing the discrete spectral efficiency problem was proposed in [15], where the discrete unit modulus constraints are directly taken into consideration in the optimization of analog precoder.

To the best of our knowledge, all the existing hybrid precoding methods are heuristic and there is no work that tries to directly solve the spectral efficiency optimization problems of various MIMO systems under hybrid precoding setups from the perspective of mathematical optimization. While some of the existing hybrid precoding methods can achieve very high spectral efficiency when the number of RF chains is strictly greater than the number of symbols and even the same performance as the fully-digital precoding method when the number of RF chains is no less than twice the number of streams, there is in some cases a significant gap between the achievable rate of the existing methods and the theoretical maximum rate. Hence, an interesting and fundamental question is: *if the spectral efficiency optimization problem is directly addressed from a perspective of mathematical optimization, is it possible for the hybrid precoding method to obtain a better system rate and how close it is to the fully-digital precoding performance when the minimum number of RF chains are used*. To answer this fundamental question, we propose an iterative hybrid precoding algorithm for multi-stream multi-user MIMO systems, aiming to directly solve the spectral efficiency optimization problem. Apparently, the main difficulty of the spectral efficiency optimization problem arises from the unit modulus constraints and the coupling among the analog precoder and the digital precoders. To address these difficulties, we apply penalty dual decomposition (PDD) method [27], [28] to the spectral efficiency optimization problem by first penalizing and dualizing the coupling constraint into the objective, and then iteratively solving the augmented Lagrangian problem using block successive upper-bound minimization (BSUM) method [29] which can naturally exploit the special structure of the unit modulus constraints. The PDD method has guaranteed convergence to KKT solutions of the spectral efficiency optimization problem. Moreover, it can be straightforwardly extended to the finite resolution phase shifter case. Our simulation results show that, even when both the transmitter and the receivers are equipped with the fewest RF chains that are required to support multi-stream transmission, hybrid precoding can still approach the performance of fully-digital precoding in both the infinite resolution phase shifter case and the finite resolution phase shifter case with several bits quantization.

*Notations*: scalars are denoted by lower-case letters, bold-face lower-case letters are used for vectors, and bold-face upper-case letters for matrices. For a matrix $\mathbf{A}$, $\mathbf{A}^T$, $\mathbf{A}^H$, $\mathbf{A}^\dagger$, $\text{Tr}(\mathbf{A})$ and $\det(\mathbf{A})$ denote its transpose, conjugate transpose, pseudo-inverse, trace, and determinant, respectively. $\mathbf{I}$ denotes an identity matrix whose dimension will be clear from the context. $|x|$, $\Re e\{x\}$ and $x^*$ are the absolute value, real part and conjugate of a complex scalar $x$, respectively, while $\|\boldsymbol{x}\|$ and $\|\mathbf{X}\|$ denote the Euclidean norm and the Frobenius norm of a complex vector $\boldsymbol{x}$ and a complex matrix $\mathbf{X}$, respectively. $\|\boldsymbol{x}\|_\infty$ denotes the infinity norm. For a matrix $\mathbf{X}$, we use $[\mathbf{X}]_{ij}$ or $\mathbf{X}(i,j)$ to denote its $(i,j)$-th entry. Particularly, we use $\mathbf{X} \in \mathcal{M}$ to denote that each entry of $\mathbf{X}$ has a unit modulus constraint, i.e., $|\mathbf{X}_{ij}| = 1$. The distribution of a circularly symmetric complex Gaussian (CSCG) random vector variable with mean $\boldsymbol{\mu}$ and covariance matrix $\mathbf{C}$ is denoted by $\mathcal{CN}(\boldsymbol{\mu}, \mathbf{C})$, and '$\sim$' stands for ' distributed as'. $\mathbb{C}^{m \times n}$ denotes the space of $m \times n$ complex matrices and $\mathbb{R}^n$ denotes the n-dimensional real vector space.

## II. SYSTEM MODEL AND PROBLEM STATEMENT

Consider a narrow band single-cell mmWave downlink multi-user multi-stream MIMO system, where a base station (BS) equipped with $N$ antennas and $N_{RF}$ ($\leq N$) transmit RF chains sends signals to $K > 1$ users, each equipped with $M > 1$ antennas and $M_{RF}$ ($\leq M$) receive RF chains. Let $d > 1$ denote the number of streams intended for each receiver. For successful symbol detection at each user, it is assumed that $d \leq \min(N_{RF}/K, M_{RF})$. Furthermore, to reduce hardware complexity, we consider a hybrid digital and analog precoding architecture for both BS and users (see Fig. 1 of [15] for an illustration of the hybrid precoding architecture).

In the multi-user hybrid precoding scheme, the BS first processes the data streams digitally at the baseband using digital precoders, and then up-converts the digitally processed signals to the carrier frequency through RF chains, followed by an analog precoder which is implemented by analog phase shifters. Let $\mathbf{V}_{RF} \in \mathbb{C}^{N \times N_{RF}}$ denote the BS analog precoder, and $\mathbf{V}_{BB_k} \in \mathbb{C}^{N_{RF} \times d}$ denote the digital precoder for user $k$'s data stream $\boldsymbol{s}_k \in \mathbb{C}^{d \times 1}$. Mathematically, the BS transmit signal after hybrid precoding is expressed as

$$\boldsymbol{x} = \mathbf{V}_{RF} \sum_{k=1}^{K} \mathbf{V}_{BB_k} \boldsymbol{s}_k \quad (1)$$

and the received signal at user $k$ is given by

$$\boldsymbol{y}_k = \mathbf{H}_k \boldsymbol{x} + \boldsymbol{n}_k \quad (2)$$

where $\mathbf{H}_k$ denotes the channel between the BS and user $k$, and $\boldsymbol{n}_k$ denotes the additive white Gaussian noise (AWGN) with zero mean and variance $\sigma^2$.

At the receiver side, user $k$ first processes the received signal by using an analog combiner $\mathbf{U}_{RF_k} \in \mathbb{C}^{M \times M_{RF}}$, and then down-converts the signals to the baseband through RF chains, followed by a digital combiner $\mathbf{U}_{BB_k} \in \mathbb{C}^{M_{RF} \times d}$ to obtain the final processed signal, given by

$$\begin{aligned}
\hat{\boldsymbol{s}}_k &= \mathbf{U}_{BB_k}^H \mathbf{U}_{RF_k}^H \boldsymbol{y}_k \\
&= \mathbf{U}_{BB_k}^H \mathbf{U}_{RF_k}^H \mathbf{H}_k \mathbf{V}_{RF} \mathbf{V}_{BB_k} \boldsymbol{s}_k \\
&+ \sum_{j \neq k}^{K} \mathbf{U}_{BB_k}^H \mathbf{U}_{RF_k}^H \mathbf{H}_k \mathbf{V}_{RF} \mathbf{V}_{BB_j} \boldsymbol{s}_j + \mathbf{U}_{BB_k}^H \mathbf{U}_{RF_k}^H \boldsymbol{n}_k.
\end{aligned} \quad (3)$$

Assuming Gaussian signaling for the data streams each with zero mean and unit variance, and that the noises $\boldsymbol{n}_k$'s and the data streams $\boldsymbol{s}_k$'s are independent of each other, the overall system spectrum efficiency maximization problem can be formulated as[1]

$$\begin{aligned}
\max_{\mathbf{V},\mathbf{U}} \sum_{k=1}^{K} &\log \det \Big( \mathbf{I} + \mathbf{U}_{RF_k}^H \mathbf{H}_k \mathbf{V}_{RF} \mathbf{V}_{BB_k} \\
&\times \mathbf{V}_{BB_k}^H \mathbf{V}_{RF}^H \mathbf{H}_k^H \mathbf{U}_{RF_k} \tilde{\boldsymbol{\Upsilon}}_k^{-1} \Big) \\
\text{s.t.} \quad &\sum_{k=1}^{K} \|\mathbf{V}_{RF} \mathbf{V}_{BB_k}\|^2 \leq P, \\
&|\mathbf{V}_{RF}(i,j)|=1, \forall i=1,2,\ldots,N, j=1,2,\ldots,N_{RF}, \\
&|\mathbf{U}_{RF_k}(i,j)|=1, \forall i=1,2,\ldots,M, j=1,2,\ldots,M_{RF}, \forall k.
\end{aligned} \quad (4)$$

where the first constraint is the BS power constraint with power budget $P$; the last two sets of unit modulus constraints are due to the fact that both the analog precoder and the analog combiners are implemented using low-cost phase shifters; and the matrix

$$\begin{aligned}
\tilde{\boldsymbol{\Upsilon}}_k &\triangleq \mathbf{U}_{RF_k}^H \Big( \sigma^2 \mathbf{I} + \sum_{j \neq k} \mathbf{H}_k \mathbf{V}_{RF} \mathbf{V}_{BB_j} \\
&\times \mathbf{V}_{BB_j}^H \mathbf{V}_{RF}^H \mathbf{H}_k^H \Big) \mathbf{U}_{RF_k}.
\end{aligned} \quad (5)$$

is the so-called interference-plus-noise covariance matrix associated with user $k$.

It was shown in [15], [18] that, when the number of transmit RF chains $N_{RF}$ are greater than or equal to twice the total number of streams, i.e., $N_{RF} \geq 2Kd$, the performance of the fully-digital precoding scheme can be perfectly achieved by the hybrid precoding scheme. This implies that problem (4) can be addressed by first solving the fully-digital precoding problem and then constructing the hybrid precoders $\mathbf{V}_{BB_k}$ and $\mathbf{V}_{RF}$ based on the fully-digital precoder. It is noted that, the fully-digital precoding problem can be addressed using the well-known WMMSE algorithm [30], [31] and the corresponding hybrid precoders can be found in closed-form; see [15, Prop. 2]. However, for the general case, the hybrid precoding problem (4) is extremely hard to solve due to not only the coupling of the digital/analog precoders but also the unit modulus constraints. To the best of our knowledge, all the existing works that deal with these difficulties are heuristic algorithms and thus inevitably incurs performance loss. Although most of heuristic algorithms for single-user MIMO scenarios perform well, it is still not known how close their performance is to the optimal one. Hence, it is important to devise an algorithm that can solve problem (4) with some theoretical guarantees (e.g., achieve at least stationary solutions to problem (4)). In this paper, we propose using penalty dual decomposition method [28] to address problem (4). Before proceeding to solving problem (4), let us first give a brief introduction to the PDD method in what follows.

---

[1]The system rate expression does not include the digital combiners explicitly since it is well-known [30] that the optimal digital combiners (i.e., MMSE receivers) can achieve the maximum system rate. In addition, for convenience, we use $\mathbf{V}$ to denote the set of variables $\mathbf{V}_{BB_k}$'s and $\mathbf{V}_{RF}$; similarly for $\mathbf{U}$ and other variables later.

## III. A BRIEF INTRODUCTION TO PDD METHOD

The PDD method is a general algorithmic framework that can be applied to the minimization of a nonconvex *nonsmooth* function subject to nonconvex coupling constraints. Considering that problem (4) has a differentiable objective function, we are here only concern about the differentiable case for ease of understanding of the PDD framework.

Consider the following problem

$$(P) \quad \min_{\boldsymbol{x}} f(\boldsymbol{x})$$
$$\text{s.t. } \boldsymbol{h}(\boldsymbol{x}) = \boldsymbol{0}, \quad (6)$$
$$\boldsymbol{x} \in \mathcal{X}.$$

where $f(\boldsymbol{x})$ is a scalar continuously differentiable function and $\boldsymbol{h}(\boldsymbol{x}) \in \mathbb{R}^{p \times 1}$ is a vector of $p$ continuously differentiable functions; the feasible set $\mathcal{X}$ is the Cartesian product of $n$ closed sets: $\mathcal{X} \triangleq \mathcal{X}_1 \times \mathcal{X}_2 \times \ldots \times \mathcal{X}_n$ with $\mathcal{X}_i \triangleq \{\boldsymbol{x}_i \mid \boldsymbol{g}_i(\boldsymbol{x}_i) \leq 0\} \subseteq \mathbb{R}^{m_i}$ and $\sum_{i=1}^{n} m_i = m$, and accordingly the optimization variable $\boldsymbol{x} \in \mathbb{R}^m$ can be decomposed as $\boldsymbol{x} = (\boldsymbol{x}_1, \boldsymbol{x}_2, \ldots, \boldsymbol{x}_n)$ with $\boldsymbol{x}_i \in \mathcal{X}_i$ $i = 1, 2, \ldots, n$; $\boldsymbol{g}_i(\boldsymbol{x}_i) \in \mathbb{R}^{q_i}$ is a vector of differentiable functions.

If no coupling constraints $\boldsymbol{h}(\boldsymbol{x}) = \boldsymbol{0}$ exist, the classical block coordination descent (BCD)-type algorithms [32] can be applied to decompose problem $(P)$ into a sequence of small-scale problems. This observation motivates us to dualize the difficult coupling constraints with appropriate penalty, and use coordinate-decomposition to perform fast computation, hence the name penalty dual decomposition method. Specifically, the PDD method applied to problem $(P)$ is a double-loop iterative algorithm. It employs inner iterations to solve to some accuracy a nonconvex augmented Lagrangian problem via an inexact or exact block coordinate descent method, while updating dual variables and a penalty parameter in outer iterations.

The PDD method is summarized in TABLE I, where the oracle 'BSUM$(P_{\varrho_k, \boldsymbol{\lambda}_k}, \tilde{\mathcal{L}}_k, \boldsymbol{z}^{k-1}, \epsilon_k)$' means that, starting from $\boldsymbol{z}^{k-1}$, the BSUM algorithm [29] or its variant randomized BSUM proposed in [28] for nonconvex constraint cases is invoked to iteratively solve problem $(P_{\varrho_k, \boldsymbol{\lambda}_k})$, given below

$$(P_{\varrho_k, \boldsymbol{\lambda}_k}) \quad \min_{\boldsymbol{x}_i \in \mathcal{X}_i} \left\{ \mathcal{L}_k(\boldsymbol{x}) \triangleq f(\boldsymbol{x}) + \boldsymbol{\lambda}_k^T \boldsymbol{h}(\boldsymbol{x}) + \frac{1}{2\varrho_k} \|\boldsymbol{h}(\boldsymbol{x})\|^2 \right\} \quad (7)$$

where $\mathcal{L}_k(\boldsymbol{x})$ is the augmented Lagrange function with dual variable $\boldsymbol{\lambda}_k$ and penalty parameter $\varrho_k$. Further, the BSUM algorithm utilizes a locally tight upper bound of $\mathcal{L}_k(\boldsymbol{x})$, denoted as $\tilde{\mathcal{L}}_k$, and it terminates when certain solution accuracy $\epsilon_k$ is reached. A basic implementation of the BSUM oracle is presented in Appendix A.

It is worthy noting that, when the penalty parameter $\varrho_k$ in (7) is sufficiently small, the term $\|\boldsymbol{h}(\boldsymbol{x})\|^2$ or $\|\boldsymbol{h}(\boldsymbol{x})\|_\infty$ will be driven to vanish after solving problem (7). That is, in this case, solving problem (7) yields a solution to problem $(P)$ satisfying the constraint $\|\boldsymbol{h}(\boldsymbol{x})\|_\infty = 0$. However, we have generally $\|\boldsymbol{h}(\boldsymbol{x})\|_\infty \neq 0$ during iterations. To measure the violation of the constraint $\boldsymbol{h}(\boldsymbol{x}) = \boldsymbol{0}$, we refer to the value of $\|\boldsymbol{h}(\boldsymbol{x})\|_\infty$ as *constraint violation*.

Define $\boldsymbol{g}(\boldsymbol{x}) \triangleq (\boldsymbol{g}_i(\boldsymbol{x}_i))_i$. Then we have the following theory regarding the convergence the PDD method, which is a direct result of [28, Theorem 3.1].

*Corollary 3.1:* Let $\{\boldsymbol{x}^k, \boldsymbol{\nu}^k\}$ be the sequence generated by Algorithm 1 for problem $(P)$, where $\boldsymbol{\nu}^k = (\boldsymbol{\nu}_i^k)_i$ denotes the Lagrange multipliers associated with the constraints $\boldsymbol{g}_i(\boldsymbol{x}_i) \leq 0, \forall i$. The stop criterion for the BSUM algorithm involved in Algorithm 1 is

$$\|\nabla_{\boldsymbol{x}} \mathcal{L}_k(\boldsymbol{x}^k) + \nabla \boldsymbol{g}(\boldsymbol{x}^k)^T \boldsymbol{\nu}^k\|_\infty \leq \epsilon_k, \quad \forall k \quad (8)$$

with $\epsilon_k, \eta_k \to 0$ as $k \to \infty$. Suppose that $\boldsymbol{x}^\star$ is a limit point of the sequence $\{\boldsymbol{x}^k\}$ and *Robinson's condition* holds for problem $(P)$ at $\boldsymbol{x}^\star$. Then $\boldsymbol{x}^\star$ is a KKT point of problem $(P)$, i.e., it satisfies the KKT condition of problem $(P)$.

A general proof of this theorem can be found in [28]. Note that Robinson's condition is a constraint qualification condition which is often used to characterize the first order optimality condition of nonconvex problems [33]–[37]. For the problem at hand, it is equivalent to the commonly used *Mangasarian-Fromovitz constraint qualification* (MFCQ) condition. Further, a simple way to set $\eta_k$ is to make it explicitly related to the constraint violation of the last iteration or the current minimum constraint violation. For example, we set $\eta_k = 0.9 \|\boldsymbol{h}(\boldsymbol{z}^{k-1})\|_\infty$ in our simulations later. Furthermore, it is reasonable to terminate the BSUM algorithm based on the progress of the objective value $\mathcal{L}_k(\boldsymbol{x}^r)$, i.e., $\frac{|\mathcal{L}_k(\boldsymbol{x}^r) - \mathcal{L}_k(\boldsymbol{x}^{r-1})|}{|\mathcal{L}_k(\boldsymbol{x}^{r-1})|} \leq \epsilon_k$. Here, the superscript '$r$' denotes the BSUM iterations. In addition, since the penalty term $\|\boldsymbol{h}(\boldsymbol{x})\|_\infty$ vanishes eventually, a practical choice of the termination condition for the PDD method is $\|\boldsymbol{h}(\boldsymbol{x}^k)\|_\infty \leq \epsilon_O$. Here, $\epsilon_O$ is some prescribed small constant, e.g., $\epsilon_O = 1e-6$. See more details in [28].

## IV. PDD METHOD FOR HYBRID PRECODING

This section proposes a PDD-based hybrid precoding method for spectral efficiency optimization. To tackle the difficulty arising from the rate function, i.e., the $\log \det(\cdot)$ function, we apply the PDD method to an equivalent problem of (4) instead of directly to (4).

### A. Equivalent formulations of (4)

Let us first rewrite problem (4) in the form of $(P)$. To make the power constraint easy to handle and remove the coupling of $\mathbf{V}_{RF}$ and $\mathbf{V}_{BB_k}$ later through penalty [cf. Eq. (13)], we

TABLE I
ALGORITHM 1: PDD METHOD FOR PROBLEM (6)

```
0.  initialize x⁰, ϱ₀ > 0, λ₀, and set 0 < c < 1, k = 1
1.  repeat
2.      xᵏ = BSUM(P_{ϱ_k,λ_k}, 𝓛̃_k, xᵏ⁻¹, ε_k)
3.      if ‖h(xᵏ)‖_∞ ≤ η_k
4.          λ_{k+1} = λ_k + (1/ϱ_k) h(xᵏ)
5.          ϱ_{k+1} = ϱ_k
6.      else
7.          λ_{k+1} = λ_k
8.          ϱ_{k+1} = cϱ_k
9.      end
10.     k = k + 1
11. until some termination criterion is met
```

introduce an auxiliary variable $\mathbf{X}_k$ such $\mathbf{X}_k = \mathbf{V}_{RF}\mathbf{V}_{BB_k}$ with the definition of

$$\mathbf{\Upsilon}_k \triangleq \mathbf{U}_{RF_k}^H \left( \sigma^2 \mathbf{I} + \sum_{j \neq k} \mathbf{H}_k \mathbf{X}_j \mathbf{X}_j^H \mathbf{H}_k^H \right) \mathbf{U}_{RF_k}. \quad (9)$$

Then we can recast problem (4) as

$$\max_{\mathbf{V},\mathbf{U},\mathbf{X}} \sum_{k=1}^K \log\det\left(\mathbf{I} + \mathbf{U}_{RF_k}^H \mathbf{H}_k \mathbf{X}_k \mathbf{X}_k^H \mathbf{H}_k \mathbf{U}_{RF_k} \mathbf{\Upsilon}_k^{-1}\right)$$
$$\text{s.t.} \sum_{k=1}^K \|\mathbf{X}_k\|^2 \leq P,$$
$$\mathbf{X}_k = \mathbf{V}_{RF}\mathbf{V}_{BB_k}, \forall k = 1,2,\ldots,K,$$
$$|\mathbf{V}_{RF}(i,j)| = 1, \forall i = 1,2,\ldots,N, j = 1,2,\ldots,N_{RF},$$
$$|\mathbf{U}_{RF_k}(i,j)| = 1, \forall i=1,2,\ldots,M, j=1,2,\ldots,M_{RF}, \forall k. \quad (10)$$

Furthermore, to address the difficulty arising from the $\log\det(\cdot)$ function, based on the theory of the well-known WMMSE method [30], we write (10) in an equivalent WMMSE form, which is shown in Proposition 4.1 with the definition of mean-square-error (MSE) matrix

$$\mathbb{E}_k(\mathbf{U},\mathbf{X}) \triangleq (\mathbf{I} - \mathbf{U}_{BB_k}^H \mathbf{U}_{RF_k}^H \mathbf{H}_k \mathbf{X}_k)(\mathbf{I} - \mathbf{U}_{BB}^H \mathbf{U}_{RF}^H \mathbf{H}_k \mathbf{X}_k)^H$$
$$+ \mathbf{U}_{BB_k}^H \mathbf{\Upsilon}_k \mathbf{U}_{BB_k}. \quad (11)$$

*Proposition 4.1:* Problem (10) is equivalent to

$$\max_{\mathbf{U},\mathbf{V},\mathbf{W},\mathbf{X}} \sum_{k=1}^K \left(\log\det(\mathbf{W}_k) - \text{Tr}(\mathbf{W}_k \mathbb{E}_k(\mathbf{U},\mathbf{X})) + d\right)$$
$$\text{s.t.} \sum_{k=1}^K \|\mathbf{X}_k\|^2 \leq P, \quad (12)$$
$$\mathbf{X}_k = \mathbf{V}_{RF}\mathbf{V}_{BB_k}, \forall k,$$
$$|\mathbf{V}_{RF}(i,j)| = 1, \forall i,j,$$
$$|\mathbf{U}_{RF_k}(i,j)| = 1, \forall i,j,k.$$

Moreover, if $\{\mathbf{U},\mathbf{V},\mathbf{W},\mathbf{X}\}$ is a KKT point of problem (12), $\{\mathbf{U},\mathbf{V},\mathbf{X}\}$ is a KKT point of problem (10), and further $\{\mathbf{U},\mathbf{V}\}$ is a KKT point of problem (4).

*Proof:* The proposition can be proven by following [30, Theorem 3] and [38, Lemma 4.1]. We omit the proof for brevity. ∎

Problem (12) is now in the form of problem $(P)$ is more tractable than problem (10) because the optimization with respect to $\mathbf{W}_k$'s is easy (see (17) below). Moreover, according to the result of Proposition 4.1, it is known that, the KKT solutions of problem (10) can be obtained by solving problem (12). Hence, in the following we solve problem (12) by using the PDD method [27].

### B. PDD method for (12)

Now we turn our attention to solving problem (12). It is readily known that the key to the PDD method is the inner iterations for solving augmented Lagrangian problems. Hence, our main efforts are devoted to developing BSUM/BCD algorithm for the augmented Lagrangian problem associated with (12). Specifically, by introducing dual variables $\mathbf{Y}_k$'s for the coupling constraints $\mathbf{X}_k = \mathbf{V}_{RF}\mathbf{V}_{BB_k}, k = 1,2,\ldots,K$, we define the augmented Lagrangian problem of (12) as follows

$$\max_{\mathbf{U},\mathbf{V},\mathbf{W},\mathbf{X}} \sum_{k=1}^K \left(\log\det(\mathbf{W}_k) - \text{Tr}(\mathbf{W}_k \mathbb{E}_k(\mathbf{U},\mathbf{X})) + d\right)$$
$$- \sum_{k=1}^K \frac{1}{2\rho} \|\mathbf{X}_k - \mathbf{V}_{RF}\mathbf{V}_{BB_k} + \rho\mathbf{Y}_k\|^2 \quad (13)$$
$$\text{s.t.} \sum_{k=1}^K \|\mathbf{X}_k\|^2 \leq P,$$
$$|\mathbf{V}_{RF}(i,j)| = 1, \forall i,j,$$
$$|\mathbf{U}_{RF_k}(i,j)| = 1, \forall i,j,k.$$

It is seen that the constraints of the above problem are separable. Hence, we use BCD-type algorithm to address (13) with the block variables $\{\mathbf{U}_{BB},\mathbf{W}\}, \mathbf{V}_{BB}, \mathbf{U}_{RF}, \mathbf{V}_{RF}$, and $\mathbf{X}$, i.e., each time we optimize one block among them while fixing the others. This leads to the following six kinds of subproblems in each BCD iteration.

*1) The subproblem w.r.t $\mathbf{U}_{BB_k}$:* The variable $\mathbf{U}_{BB_k}$ is updated by minimizing the term $\text{Tr}(\mathbf{W}_k \mathbb{E}_k(\mathbf{U},\mathbf{X}))$, i.e., solving

$$\min_{\mathbf{U}_{BB_k}} \text{Tr}(\mathbf{W}_k \mathbf{U}_{BB_k}^H \mathbf{U}_{RF_k}^H \mathbf{A}_k \mathbf{U}_{RF_k} \mathbf{U}_{BB_k})$$
$$- 2\Re e\left\{\text{Tr}(\mathbf{W}_k \mathbf{U}_{BB_k}^H \mathbf{U}_{RF_k}^H \mathbf{H}_k \mathbf{X}_k)\right\} \quad (14)$$

where

$$\mathbf{A}_k \triangleq \sigma_k^2 \mathbf{I} + \sum_{j=1}^K \mathbf{H}_k \mathbf{X}_j \mathbf{X}_j^H \mathbf{H}_k^H. \quad (15)$$

Since each matrix $\mathbf{W}_k$ is positive definite, the optimal solution $\mathbf{U}_{BB_k}$ to problem (14) takes the form

$$\mathbf{U}_{BB_k} = \left(\mathbf{U}_{RF_k}^H \mathbf{A}_k \mathbf{U}_{RF_k}\right)^\dagger \left(\mathbf{U}_{RF_k}^H \mathbf{H}_k \mathbf{X}_k\right). \quad (16)$$

The reason why we use pseudo-inverse here is that the matrix $\mathbf{U}_{RF_k}^H \mathbf{A}_k \mathbf{U}_{RF_k}$ may be rank-deficient during iterations.

*2) The subproblem w.r.t $\mathbf{W}_k$:* Consequently, given the optimal $\mathbf{U}_{BB_k}$ in (16), the optimal $\mathbf{W}_k$ can be expressed as

$$\mathbf{W}_k = \arg\min_{\bar{\mathbf{W}}} \log\det(\bar{\mathbf{W}}_k) - \text{Tr}(\bar{\mathbf{W}}_k \mathbb{E}_k(\mathbf{U},\mathbf{X}))$$
$$= \mathbb{E}_k(\mathbf{U},\mathbf{X})^{-1}$$
$$= \left(\mathbf{I} - \mathbf{U}_{BB_k}^H \left(\mathbf{U}_{RF_k}^H \mathbf{H}_k \mathbf{X}_k\right)\right)^{-1}. \quad (17)$$

where the last equality is obtained by plugging (16) into $\mathbb{E}_k(\mathbf{U},\mathbf{X})$. Moreover, it can be verified that the bracketed matrix in the last equality is positive definite and so is the matrix $\mathbf{W}_k$.

*3) The subproblem w.r.t $\mathbf{V}_{BB_k}$:* The variable $\mathbf{V}_{BB_k}$ appears only in the summand of the second term of the objective of (13), i.e., $\|\mathbf{X}_k - \mathbf{V}_{RF}\mathbf{V}_{BB_k} + \rho\mathbf{Y}_k\|^2$. Thus, the subproblem with respect to $\mathbf{V}_{BB_k}$ is given by

$$\min_{\mathbf{V}_{BB_k}} \|\mathbf{V}_{RF}\mathbf{V}_{BB_k} - \mathbf{Z}_k\|^2 \quad (18)$$

where $\mathbf{Z}_k \triangleq \mathbf{X}_k + \rho\mathbf{Y}_k$. Solving the above unconstrained quadratic optimization problem, we obtain the optimal $\mathbf{V}_{BB_k}$ as follows

$$\mathbf{V}_{BB_k} = (\mathbf{V}_{RF})^\dagger \mathbf{Z}_k. \quad (19)$$

*4) The subproblem w.r.t $\mathbf{X}$:* The variables $\mathbf{X}_k$'s appear in both the first and the second terms of the objective of (13). Thus, the subproblem with respect to $\mathbf{X}$ is given by

$$\min_{\mathbf{X}} \sum_{k=1}^K \text{Tr}\left(\mathbf{W}_k \mathbb{E}_k(\mathbf{U},\mathbf{X})\right) + \sum_{k=1}^K \frac{1}{2\rho} \|\mathbf{X}_k + \rho\mathbf{Y}_k - \mathbf{V}_{RF}\mathbf{V}_{BB_k}\|^2$$
$$\text{s.t.} \sum_{k=1}^K \|\mathbf{X}_k\|^2 \leq P. \quad (20)$$

For notational convenience, we define

$$\mathbf{A}_\rho \triangleq \sum_{j=1}^K \left(\mathbf{H}_j^H \mathbf{U}_{RF_j} \mathbf{U}_{BB_j} \mathbf{W}_j \mathbf{U}_{BB_j}^H \mathbf{U}_{RF_j}^H \mathbf{H}_j\right) + \frac{1}{2\rho}\mathbf{I},$$
$$\mathbf{B}_{\rho,k} \triangleq \mathbf{H}_k^H \mathbf{U}_{RF_k} \mathbf{U}_{BB_k} \mathbf{W}_k + \frac{1}{2}\left(\frac{1}{\rho}\mathbf{V}_{RF_k}\mathbf{V}_{BB_k} - \mathbf{Y}_k\right).$$

Then, with some appropriate rearrange, problem (20) can be equivalently written as

$$\min_{\mathbf{X}} \sum_{k=1}^K \left(\text{Tr}(\mathbf{X}_k^H \mathbf{A}_\rho \mathbf{X}_k) - 2\Re e\left\{\text{Tr}(\mathbf{X}_k^H \mathbf{B}_{\rho,k})\right\}\right)$$
$$\text{s.t.} \sum_{k=1}^K \|\mathbf{X}_k\|^2 \leq P. \quad (21)$$

Further, by introducing a Lagrange multiplier $\mu \geq 0$ to the power constraint, we can express the optimal $\mathbf{X}_k$ as

$$\mathbf{X}_k = (\mathbf{A}_\rho + \mu\mathbf{I})^{-1} \mathbf{B}_{\rho,k} \quad (22)$$

where the optimal multiplier $\mu$ can be easily found by using Bisection method such that

$$\sum_{k=1}^K \text{Tr}\left(\mathbf{B}_{\rho,k}^H \left(\mathbf{A}_\rho + \mu\mathbf{I}\right)^{-2} \mathbf{B}_{\rho,k}\right) = P,$$

equivalently,

$$\sum_{k=1}^K \sum_{i=1}^N \frac{b_{k,i}}{(a_i+\mu)^2} = P$$

where $b_{k,i} = [\mathbf{U}_\rho \mathbf{B}_{\rho,k} \mathbf{B}_{\rho,k}^H \mathbf{U}_\rho^H]_{ii}$, $\mathbf{U}_\rho$ is a unitary matrix consisting of the eigenvectors of $\mathbf{A}_\rho$ and $a_i$'s are the corresponding eigenvalues, i.e., $\mathbf{U}_\rho \text{diag}\{a_1, a_2, \ldots, a_N\} \mathbf{U}_\rho^H$ is the eigenvalue decomposition of $\mathbf{A}_\rho$.

*5) The subproblem w.r.t $\mathbf{V}_{RF}$:* The variable $\mathbf{V}_{RF}$ only appears in the summand of the second term of the objective of (13). Thus, the subproblem with respect to $\mathbf{V}_{RF}$ is given by

$$\min_{\mathbf{V}_{RF}\in\mathcal{M}} \sum_{k=1}^K \|\mathbf{X}_k - \mathbf{V}_{RF}\mathbf{V}_{BB_k} + \rho\mathbf{Y}_k\|^2 \quad (23)$$

By appropriate rearrange, the above problem can be equivalently formulated as follows

$$\min_{\mathbf{V}_{RF}\in\mathcal{M}} \text{Tr}(\mathbf{V}_{RF}^H \mathbf{V}_{RF} \mathbf{C}_{V_{RF}}) - 2\Re e\left\{\text{Tr}(\mathbf{V}_{RF}^H \mathbf{B}_{V_{RF}})\right\} \quad (24)$$

where $\mathbf{C}_{V_{RF}} \triangleq \sum_{k=1}^K \mathbf{V}_{BB_k}\mathbf{V}_{BB_k}^H$, $\mathbf{B}_{V_{RF}} \triangleq \sum_{k=1}^K (\mathbf{X}_k + \rho\mathbf{Y}_k)\mathbf{V}_{BB_k}^H$. Since the unit modulus constraints are separable, we can use one-iteration BCD-type algorithm to recursively solve problem (24). See the details in Appendix B.

*6) The subproblem w.r.t $\mathbf{U}_{RF_k}$:* The variable $\mathbf{U}_{RF_k}$ appears only in the term $\mathbb{E}_k(\mathbf{U},\mathbf{X})$. Thus, the subproblem with respect to $\mathbf{U}_{RF_k}$ is given by

$$\min_{\mathbf{U}_{RF_k}\in\mathcal{M}} \text{Tr}(\mathbf{W}_k \mathbb{E}_k(\mathbf{U},\mathbf{X})) \quad (25)$$

By appropriate rearrange, the above problem can be equivalently formulated as follows

$$\min_{\mathbf{U}_{RF_k}\in\mathcal{M}} \text{Tr}\left(\mathbf{U}_{RF_k}^H \mathbf{A}_k \mathbf{U}_{RF_k} \mathbf{C}_{U_{RF_k}}\right) - 2\Re e\left\{\text{Tr}(\mathbf{U}_{RF_k}^H \mathbf{B}_{U_{RF_k}})\right\} \quad (26)$$

where $\mathbf{A}_k$ is defined in (15), $\mathbf{C}_{U_{RF_k}} \triangleq \mathbf{U}_{BB_k}\mathbf{W}_k \mathbf{U}_{BB_k}^H$, $\mathbf{B}_{U_{RF_k}} \triangleq \mathbf{H}_k \mathbf{X}_k \mathbf{W}_k \mathbf{U}_{BB_k}^H$. Again, we can use one-iteration BCD-type algorithm to address problem (26). See the details in Appendix B.

The BCD-type algorithm for (13) is summarized in TABLE II. Note that, all updates can be implemented in parallel for different users. For instance, once $\mathbf{U}_{RF_k}$'s and $\mathbf{X}_k$'s are given, each $\mathbf{U}_{BB_k}$ can be updated independently. By using parallel implementation, the proposed BCD-type algorithm is scalable to the number of users. In particular, when $N >> M$, it can be shown that the complexity of Algorithm 2 is dominated by Step 7, which is $O(N^3 K I_2)$. Here, $I_2$ denotes the maximum number of iterations required by Algorithm 2.

In each iteration of the PDD method, after running Algorithm 2, we update the dual variable $\mathbf{Y}_k$ or the penalty parameter $\rho$ according to the constraint violation condition $\max_k \|\mathbf{X}_k - \mathbf{V}_{RF}\mathbf{V}_{BB_k}\|_\infty \leq \eta_k$ (see Step 3 in TABLE I). That is, if the constraint violation condition is satisfied, we update the dual variables $\mathbf{Y}_k$'s according to $\mathbf{Y}_k \leftarrow \mathbf{Y}_k + \frac{1}{\rho}(\mathbf{X}_k -$

$\mathbf{V}_{RF}\mathbf{V}_{BB_k}$), $\forall k$; otherwise, we increase the penalty parameter by updating $\rho = c\rho$ where $0 < c < 1$. The PDD method is terminated when the feasibility of the penalized constraints (i.e., $\mathbf{X}_k = \mathbf{V}_{RF}\mathbf{V}_{BB_k}, \forall k$) is approximately achieved; see the discussion in the end of Section III. In addition, it can be shown[2] that the MFCQ condition is satisfied for problem (12) at any feasible solution $(\{\mathbf{D}_{\mathbf{X}_k}\}, \mathbf{D}_{\mathbf{V}_{RF}}, \mathbf{D}_{\mathbf{U}_{RF}})$. Hence, we can conclude that the PDD method converges to the set of KKT solutions of problem (4) according to the results of Proposition 4.1 and Corollary 3.1.

TABLE II
ALGORITHM 2: BCD-TYPE ALGORITHM FOR PROBLEM (13)

0. initialize $\mathbf{V}_{RF}, \mathbf{V}_{BB_k}, \mathbf{U}_{RF_k}, \mathbf{U}_{BB_k}, \forall k$, such the constraints and set $\mathbf{X}_k = \mathbf{V}_{RF}\mathbf{V}_{BB_k}, \forall k$
1. **repeat**
2.   update $\mathbf{U}_{BB_k}$'s according to (16)
3.   update $\mathbf{W}_k$'s according to (17)
4.   update $\mathbf{V}_{RF}$ by solving (24) using Algorithm 4
5.   update $\mathbf{U}_{RF_k}$'s by solving (26) using Algorithm 4
6.   update $\mathbf{V}_{BB_k}$'s according to (19)
7.   update $\mathbf{X}_k$'s according to (22) using Bisection method
8. **until** some termination criterion is met

## V. MATRIX-APPROXIMATION-BASED HYBRID PRECODING METHOD

The PDD method requires running Algorithm 2 repeatedly. In this section, we provide a lightweight solution based on **m**atrix **ap**proximation (MAP).

The intuition behind MAP is that [6], the performance achieved by hybrid precoding is upper bounded by the one achieved by the fully-digital precoding. Hence, if the optimal fully-digital precoder/decoder can be approximated well using hybrid precoder/decoder, we can obtain a good spectral efficiency performance that is close to that of the fully-digital precoding. Mathematically, given the optimal fully-digital precoder $\mathbf{V}_{opt_k}, \forall k$, MAP for precoder design[3] is to find a structured matrix $\mathbf{V}_k = \mathbf{V}_{RF}\mathbf{V}_{BB_k}$ to approximate $\mathbf{V}_{opt,k}, \forall k$, i.e., solving the following matrix approximation problem

$$\min_{\mathbf{V}_{RF}, \mathbf{V}_{BB}} \|\mathbf{V}_{opt} - \mathbf{V}_{RF}\mathbf{V}_{BB}\|^2 \quad (27)$$
$$\text{s.t.} \ \mathbf{V}_{RF} \in \mathcal{M}$$

where $\mathbf{V}_{opt} \triangleq [\mathbf{V}_{opt_1} \ \mathbf{V}_{opt_2} \ \ldots \ \mathbf{V}_{opt_K}]$, $\mathbf{V}_{BB} \triangleq [\mathbf{V}_{BB_1} \ \mathbf{V}_{BB_2} \ \ldots \ \mathbf{V}_{BB_K}]$ and $\mathcal{M} \triangleq \{\mathbf{V}_{RF} \mid |\mathbf{V}_{RF}(i,j)| = 1, \forall i,j\}$. Note that, for tractability, we here have neglected the coupling power constraint which will be taken into consideration in the end by scaling $\mathbf{V}_{BB_k}$'s to satisfy the power constraint. However, even after removing this difficulty, the problem is still difficult to solve due to the unit modulus constraint.

[2]The MFCQ for this problem is equivalent to verify that 1) the equality constraint gradients are linearly independent, and 2) there exists $(\{\mathbf{D}_{\mathbf{X}_k}\}, \mathbf{D}_{\mathbf{V}_{RF}}, \mathbf{D}_{\mathbf{U}_{RF}})$ such that $\mathbf{D}_{\mathbf{X}_k} - \mathbf{V}_{RF}\mathbf{D}_{\mathbf{U}_{RF}} - \mathbf{D}_{RF}\mathbf{U}_{RF} = 0, \forall k$ and $\sum_{k=1}^K \text{Tr}(\mathbf{X}_k^H \mathbf{D}_{\mathbf{X}_k}) < 0$. Note that both are easy to be verified.
[3]The decoder can be designed in a similar way.

The quadratic nature of the objective function motivates us to use BCD-type algorithm to address problem (27). Specifically, we divide the variable $(\mathbf{V}_{RF}, \mathbf{V}_{BB})$ into $N_{RF}d + 1$ blocks, i.e., $\mathbf{V}_{BB}, \mathbf{V}_{RF}(i,j), i = 1,2,\ldots,N_{RF}, j = 1,2,\ldots,d$, and successively update each block by solving (27) for the selected block while fixing the others. Clearly, given $\mathbf{V}_{RF}$, $\mathbf{V}_{BB}$ can be updated in closed-form: $\mathbf{V}_{BB} = \mathbf{V}_{RF}^\dagger \mathbf{V}_{opt}$. Further, according to Appendix B, each entry of $\mathbf{V}_{RF}$, $\mathbf{V}_{RF}(i,j)$, can be also updated in closed-form with the other variables fixed. Note that the proposed algorithm for problem (27) differs significantly from the alternating optimization method proposed in [7], where the analog precoder $\mathbf{V}_{RF}$ is updated as a whole by using manifold optimization (MO) method [39]. Since the MO method is a gradient-descent-like iterative algorithm, our algorithm is more efficient than the alternating optimization method in [7] by using closed-form update of $\mathbf{V}_{RF}$. This will be verified later using numerical examples (see Fig. 3).

To summarize, the MAP method is a two-phase hybrid precoding method. In the first phase, we obtain the fully-digital precoder/decoder using the well-known WMMSE algorithm. In the second phase, we run BCD method to find the hybrid precoders/decoders. Assuming the number of transmit antennas $N$ is much larger than the number of user receive antennas $M$, it can be shown that the complexity of the first phase is $O(N^3 K I_{wmmse})$ where $I_{wmmse}$ denotes the number of iterations required by the WMMSE method. Furthermore, as shown in Appendix B, updating all the entries of $\mathbf{V}_{RF}$ once requires complexity of $O(N^2 N_{RF}^2)$. Similarly, updating $\mathbf{U}_{RF_k}$'s requires complexity of $O(KM^2 M_{RF}^2)$. Hence, the complexity of the second phase is $O(I_{bcd}(N^2 N_{RF}^2 + KM^2 M_{RF}^2))$ where $I_{bcd}$ denotes the maximum number of iterations required for updating precoders/decoders. Therefore, the total complexity of the MAP method is cubic in the number of BS antennas. It is lower than that of the PDD method because the latter requires repeatedly running Algorithm 2 which also has cubic complexity in the number of BS antennas.

## VI. HYBRID PRECODING WITH FINITE RESOLUTION PHASE SHIFTERS

So far, we have assumed that arbitrary resolution phase shifters are available for realizing analog precoders. However, the hardware for accurate phase control could be very expensive. Furthermore, infinite resolution phase shifter is not always practical for large-array systems. Hence, we need to consider hybrid precoding with finite resolution phase shifters, i.e., the candidate phases for each phase shifter is finite.

Let $\mathcal{F}$ denote the set of finite phases, with $|\mathcal{F}| = 2^b$, where $b$ is the number of bits used to quantize the phases. The corresponding spectral efficiency optimization problem turns

out to be

$$\max_{\mathbf{V},\mathbf{U}} \sum_{k=1}^{K} \log \det \left( \mathbf{I} + \mathbf{U}_{BB_k}^H \mathbf{U}_{RF_k}^H \mathbf{H}_k \mathbf{V}_{RF} \mathbf{V}_{BB_k} \right.$$
$$\left. \times \mathbf{V}_{BB_k}^H \mathbf{V}_{RF}^H \mathbf{H}_k^H \mathbf{U}_{RF_k} \mathbf{U}_{BB_k} \mathbf{\Upsilon}_k^{-1} \right) \quad (28)$$
$$\text{s.t.} \sum_{k=1}^{K} \|\mathbf{V}_{RF} \mathbf{V}_{BB_k}\|^2 \leq P,$$
$$\mathbf{V}_{RF}(i,j) \in \mathcal{F}, \forall i,j$$
$$\mathbf{U}_{RF}(i,j) \in \mathcal{F}, \forall i,j,k$$

Due to the combinatorial nature of the phases available for the phase shifters, problem (28) generally requires exhaustive search which however is computationally prohibitive in practice. A heuristic way to deal with the constraints of finite phases is to first addressing the spectral efficiency optimization problem under the assumption of infinite resolution phase and then quantize each component of the analog precoders to the nearest point in the set $\mathcal{F}$. This heuristic method could work effectively when the number of available phases is large (i.e., relatively high resolution phase). However, for the low resolution case (e.g., $b \leq 4$), the effect of phase quantization could be significant. Hence, it is important to devise algorithms that can incorporate the constraints of finite resolution phases directly into the optimization procedure.

Fortunately, our PDD algorithm[4] proposed above can be easily adapted to the finite resolution phase case. The modification lies only in Step 4 of Algorithm 4, i.e., modify Step 4 as follows

$$x = \arg \min_{\mathbf{X}(i,j) \in \mathcal{F}} \Re e \left\{ b^* \mathbf{X}(i,j) \right\}$$

The above problem can be globally solved via one-dimensional exhaustive search. As a result, the modified Algorithm 4 requires complexity of $O(I_3 mn(mn + 2^b))$.

Finally, we make a remark on iterations of our algorithms. All the algorithms proposed for both the infinite resolution phase case and the finite resolution phase case are iterative algorithms. While the algorithms requires a number of iterations for achieving convergence, we can terminate them early at the price of system spectral efficiency performance. It is worth mentioning that, when the algorithms are terminated early, we need to scale $\mathbf{V}_{BB_k}$'s such the BS power constraint and eventually obtain a feasible solution to the spectral efficiency optimization problem.

## VII. SIMULATION RESULTS

This section presents simulation results to illustrate the performance of the proposed hybrid precoding algorithms. Since we focus on hybrid precoding for *multi-stream multi-user* MIMO cases without exploiting the angle information of channel knowledge, we first present the simulation results for MU-MIMO systems and then for SU-MIMO and multi-user MISO systems as special cases. The recent work [18] has shown that the regularized block diagonalization (RBD)-based hybrid precoding method performs better than ZF-based, BD-based, and MMSE-based hybrid precoding methods for multi-user MIMO cases. Hence, in the simulations, we compare our hybrid precoding methods with the RBD method in addition to the benchmark performance—the fully digital precoding schemes in terms of the achieved spectral efficiency[5].

As in [15], we use a geometric channel model of $L = 15$ paths with uniform linear array antenna configurations. Specifically, the channel matrix between the BS and each user is expressed as

$$\mathbf{H}_k = \sqrt{\frac{NM}{L}} \sum_{\ell=1}^{L} \alpha_k^\ell \boldsymbol{a}_r(\phi_k^\ell) \boldsymbol{a}_t(\varphi_k^\ell)^H, \forall k. \quad (29)$$

where $\alpha_k^\ell \sim \mathcal{CN}(0,1)$ is the complex gain of the $\ell$-th path, $\boldsymbol{a}_r(\phi_k^\ell)$ and $\boldsymbol{a}_t(\varphi_k^\ell)$ are respectively the *normalized* receive and transmit array response vector at the azimuth angle of $\phi_k^\ell \in [0, 2\pi)$ and $\varphi_k^\ell \in [0, 2\pi)$, which are given by

$$\boldsymbol{a}(\theta) = \frac{1}{N} \left[ 1, e^{j\pi \sin(\theta)}, \ldots, e^{j(N-1)\pi \sin(\theta)} \right]^T. \quad (30)$$

In our simulations, it is assumed that the base station is equipped with $N = 64$ antennas and each user with $M = 16$ antennas. Unless otherwise specified, we set $N_{RF} = 8$ and $M_{RF} = 4$. For the PDD method, the initial penalty parameter $\rho$ is set to $100/N$ and the control parameter $c$ is set to $0.8$. Furthermore, we set $\eta_0 = \epsilon_0 = 1e - 3$ and $\epsilon_k = c\epsilon_{k-1}$. Moreover, in practical implementation, we set the maximum number of inner BSUM/BCD iterations of the PDD method[6] to 30. The simulation results versus $SNR$ are averaged over 100 channel realizations, where $SNR$ is defined by $SNR = 10 \log_{10}(\frac{P}{\sigma^2})$.

### A. Convergence performance of the PDD method

In the first set of simulations, we examine the convergence performance of the proposed methods. To obtain a benchmark performance, we set $K = d = 2$ so that we have $N_{RF} = 2Kd$ and $M_{RF} = 2d$. For this setup, it is known that the fully-digital precoding performance can be perfectly achieved by the hybrid precoding in the infinite resolution phase shifter case [15].

First, we examine the convergence performance of the PDD method. For 100 randomly generated multi-user channels, we run the PDD method for each channel with different random initialization. Note that, different problem instances yield different spectral efficiency value. To remove this effect and also clearly demonstrate how much percentage of fully-digital precoding performance the hybrid precoding can achieve, we normalize the objective value of problem (12) by the spectral efficiency value of the fully-digital precoding, and plot them in the left subfigure of Figs. 1 and 2, where the average convergence behavior of the PDD method is shown for the case of infinite resolution phase shifters (denoted by $b = \infty$ for short) and finite resolution phase shifters with $b = 1$, respectively. In terms of the objective value and the constraint violation, it is observed from the plots that the PDD method can converge well within 50 iterations for both cases. Furthermore, it is seen that the hybrid precoding can achieve the same performance as the fully-digital precoding in the example of infinite resolution phase shifter case, implying its excellent convergence performance in achieving possibly optimal solutions. While for the finite resolution phase shifter case with $b = 1$, the hybrid precoding can achieve about 73.7 percent of the fully-digital precoding performance in this example.

---

[4]It is worthy mentioning that, as $\mathcal{F}$ is a discrete set, the convergence result in Corollary 3.1 does not apply to the finite resolution case. While numerical results still show good convergence performance for the PDD method in the finite resolution case, the convergence issue remains open. However, our method can serve as a reference point for studying the performance of hybrid precoding architecture with finite resolution phase shifters in comparison with fully digital precoding.

[5]The most costly step of the RBD method is the SVD operations performed on the user channels (each with size of $N \times M$) and the analog precoder (with size of $N \times N_{RF}$). Hence, the RBD method has quadratic complexity in the number of BS antennas, which is lower than the cubic complexity of the PDD method.

[6]When the constraint violation is satisfactory, the maximum iteration strategy can be used to avoid the possibly slow convergence of the B-SUM/BCD algorithm due to large penalty (i.e., $1/\rho$ is large). From our numerical experience, this strategy works effectively without sacrificing any performance.

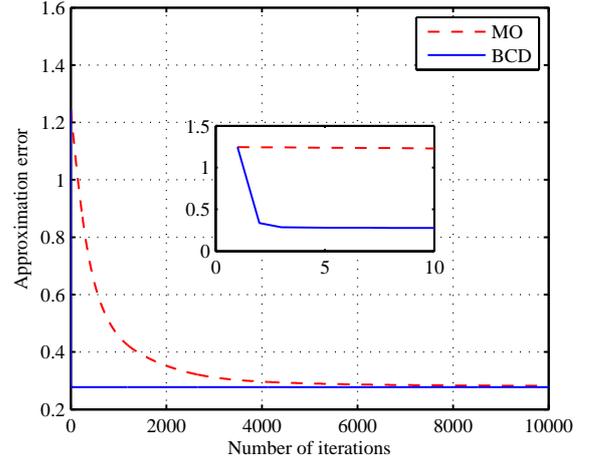

Fig. 3. The convergence performance of BCD (for (27) with some fixed $\mathbf{V}_{BB}$) versus MO.

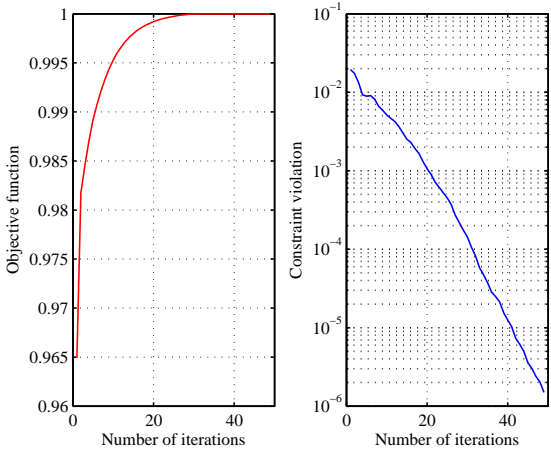

Fig. 1. The average convergence behavior of PDD with $b = \infty$.

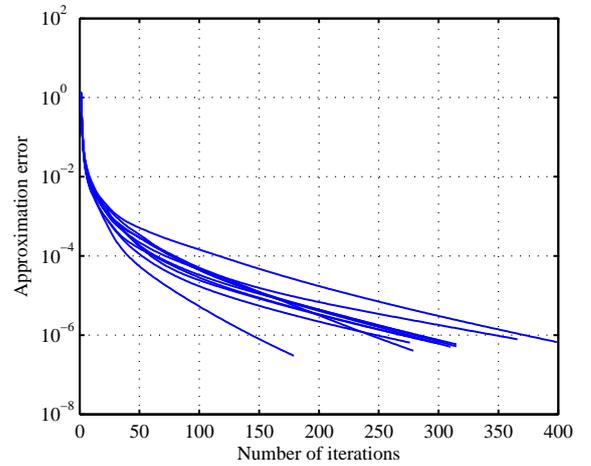

Fig. 4. Ten examples of convergence behavior of BCD (for (27)) with random initialization. Each curve corresponds to an example of convergence.

method in [7]. Since the two methods have the same way in the update of $\mathbf{V}_{BB}$ but totally different ways in the update of $\mathbf{V}_{RF}$, we here focus on the comparison of the efficiency of the update of $\mathbf{V}_{RF}$ and thus first compare the convergence performance of the BCD method and the MO method when they are applied to problem (27) with fixed $\mathbf{V}_{BB}$. Figure 3 illustrates an average convergence behavior of the two methods over ten problem instances, where 'Approximation error' denotes the objective value of problem (27). It is seen that the BCD method requires significantly less iterations (generally serveral iterations) for convergence than the MO method, while both can achieve the same convergence result. Furthermore, Figure 4 shows that, with different initializations, the MAP method can always achieve global optimality of problem (27). This will be further verified by the simulation results later.

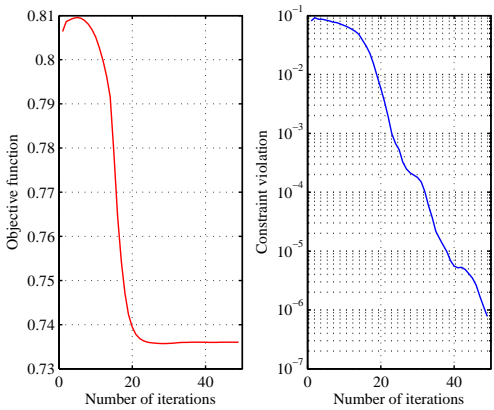

Fig. 2. The average convergence behavior of PDD with $b = 1$.

Second, we examine the convergence performance of the MAP method as compared to the alternating optimization

### B. Hybrid precoding with infinite resolution phase shiters

In the second set of simulations, we demonstrate the spectral efficiency performance of the proposed hybrid precoding methods as compared to the benchmark fully-digital precoding performance and the existing multi-user hybrid precoding method—RBD [18] in the infinite resolution phase shifter case. Figures 5 shows the spectral efficiency performance of various precoding methods for the case of $K = d = 2$. In this case, it is known that the fully-digital precoding performance is achievable by the hybrid precoding. Figure 5 shows that both the PDD method and the MAP method can achieve the same performance as the fully-digital precoding. This again verifies the excellent convergence performance of the PDD method and the MAP method (possibly achieve global optimality in this case). Moreover, it is observed from Fig. 5 that both the PDD method and the MAP method has better performance than the RBD method with about 1dB gain.

Figures 6 shows the spectral efficiency performance of various hybrid precoding methods for the cases when $N_{RF} = Kd < 2Kd$. It is seen that, even for the case when $N_{RF} < 2Kd$, the PDD method can achieve a performance that is extremely close to the FD precoding performance. Furthermore, one can see that, unlike the case of $N_{RF} \geq 2Kd$ (where the MAP method has the same performance as the PDD method), the PDD method has a better performance than the MAP method in the cases of $N_{RF} < 2Kd$. Moreover, it is observed again that both the PDD method and the MAP method outperform the RBD method, and the performance gaps increase with the SNR. The reason for this observation is explained as follows. Recall that both the MAP method and the RBD method first neglects the power constraint and then scale the digital precoder $\mathbf{V}_{BB_k}$'s to satisfy the power constraint. Such a heuristic scaling approach inevitably incurs performance degradation especially when $P$ is large (i.e., the SNR is large). Particularly, when $SNR = 6$, the PDD method improves the spectral efficiency of the MAP method and the RBD method by 5 bps/Hz and 8 bps/Hz, respectively.

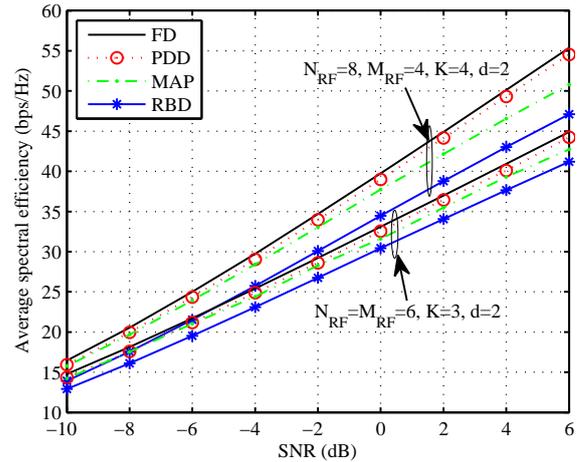

Fig. 6. Spectral efficiency achieved by different methods versus SNR when $K = 4$ and $d = 2$.

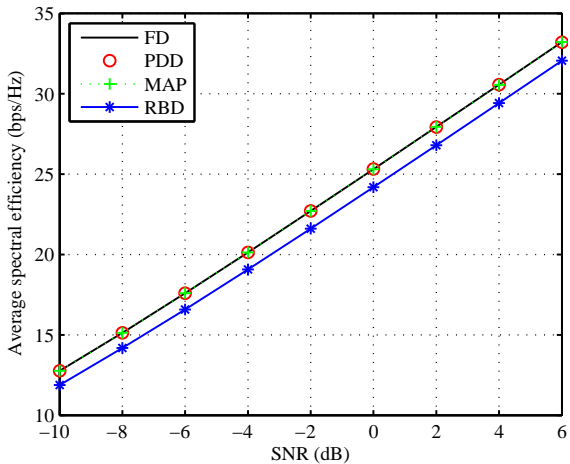

Fig. 5. Spectral efficiency achieved by different methods versus SNR when $N_{RF} = 8$, $M_{RF} = 4$, and $K = d = 2$.

### C. Hybrid precoding with finite resolution phase shiters

In the third set of simulations, we demonstrate the spectral efficiency performance achieved by the PDD-based hybrid precoding method in the finite resolution phase shifter case. Since it is not possible to get good matrix approximations of the fully-digital precoder/decoders in this case, the MAP method does not work well and thus its simulation result is not shown here. Furthermore, since the RBD method was proposed only for the infinite resolution phase shifter case in [18], we adapt it to the finite resolution phase shifter case by quantizing the analog precoder/decoders[7] followed by the design of the digital precoder/decoders. For convenience, we still refer to the modified RBD method in the finite resolution phase shifter case as 'RBD' in the plot. In addition, since the hybrid precoding problem has much more local maxima in the finite resolution phase shifter case than in the infinite resolution phase shifter case, the PDD method could get easily trapped in some bad point and thus may result in bad performance. To deal with this issue, we run 20 iterations of the PDD method assuming infinite resolution phase shifters to get a good initialization, followed by the PDD method of the finite resolution phase shifter case[8]. Such a strategy makes the PDD method work very well in the finite resolution phase shifter case from our numerical experience.

Figure 7 illustrates the spectral efficiency of the PDD method and the modified RBD method when three bits quantization is used, i.e., $b = 3$. It is observed that the PDD method can still achieve most of the spectral efficiency of the FD precoding in this example of finite resolution phase shifter case. Moreover, one can see again that the PDD method outperforms the RBD method and their gap increases with the SNR. For example, the gap between the PDD method and the modified RBD method is about 5 bps/Hz when $SNR = 6$ dB, while it is much smaller (almost negligible) when $SNR = -10$ dB.

Figure 8 illustrates the spectral efficiency of the PDD method versus SNR when different quantization levels are used. It is observed that, the spectral efficiency of the PDD method improves when more bits are used in the phase quantization, and the improvement shrinks as the quantization level increases. Particularly, one can see that three bits quantization is enough for achieving about 95% of the fully-digital precoding performance in this example.

---

[7]Specifically, we quantize each element of the analog precoder/decoder to the nearest point in the set $\mathcal{F}$.

[8]Note that the penalty parameter $\rho$ needs not to be re-initialized. Thus, it is in essence running PDD once.

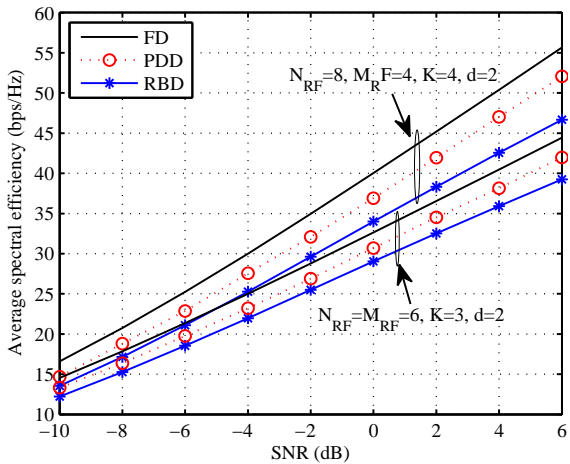

Fig. 7. Spectral efficiency achieved by different methods versus SNR when $b = 3$.

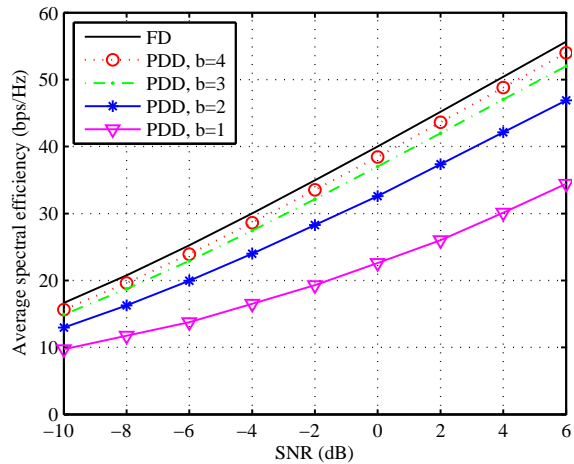

Fig. 8. Spectral efficiency achieved by PDD versus SNR for different quantization levels when $K = 4$ and $d = 2$.

Figure 9 shows the spectral efficiency of the PDD method when different number of RF chains are used with the lowest resolution phase shifters, i.e., $b = 1$. It is seen that the performance gap between the PDD-based hybrid precoding and the FD precoding can be reduced by increasing the number of RF chains. Therefore, the number of RF chains can be used to trade off the resolution of the phase shifters in hybrid precoding design.

Lastly, we consider the cases when the numbers of transmit/receive RF chains fullfil the minimum requirement, i.e., $N_{RF} = Kd$ and $M_{RF} = d$, and examine how much percentage of the FD precoding performance the hybrid precoding can achieve. In the simulations, we test five system setups with different combination of $K$ and $d$ which satisfies $N_{RF} = Kd$ and $M_{RF} = d$. For each setup, we run the PDD method for 100 randomly generated channels under three quantization levels $b = \infty$, $b = 4$, and $b = 2$. The minimum, average, maximum spectral efficiency of the PDD method relative to the spectral efficiency of the FD precoding are respectively listed in TABLE I for $SNR=0$ dB. It can be observed that, in the infinite resolution phase shifter case, the hybrid precoding can achieve more than 95 percentage of the FD precoding performance, implying the excellent efficiency of the hybrid precoding while enjoying the benefit of significantly reducing the number of RF chains in Massive MIMO systems. Furthermore, even with low resolution phase shifters, the hybrid precoding can still achieve most percentage of the FD precoding performance, e.g., it is on average about 95% when $b = 4$ and about 80% when $b = 2$.

### D. Hybrid precoding for special cases: SU-MIMO & MU-MISO

In this set of simulations, we show the hybrid precoding performance achieved by the PDD method for SU-MIMO

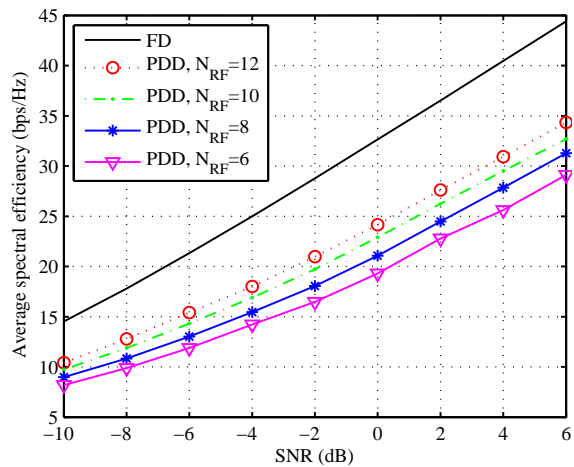

Fig. 9. Spectral efficiency achieved by PDD versus SNR for different $N_{RF}$ when $M_{RF} = 4$, $K = 3$, $d = 2$ and $b = 1$.

TABLE III
THE ACHIEVABLE RELATIVE PERFORMANCE OF HYBRID PRECODING AS COMPARED TO FULLY-DIGITAL PRECODING ($N_{RF} = Kd, M_{RF} = d$)

| Value | $(K, d)$ | $(2, 2)$ | $(4, 2)$ | $(2, 4)$ | $(2, 3)$ | $(3, 2)$ |
|---|---|---|---|---|---|---|
| $b = \infty$ | min. | 94.35% | 94.74% | 94.54% | 95.00% | 93.46% |
|  | avg. | 96.84% | 96.86% | 97.31% | 97.12% | 96.85% |
|  | max. | 98.78% | 98.10% | 98.23% | 99.19% | 98.50% |
| $b = 4$ | min. | 92.10% | 85.00% | 91.20% | 91.70% | 92.67% |
|  | avg. | 95.21% | 94.80% | 94.57% | 94.97% | 95.10% |
|  | max. | 97.30% | 97.11% | 96.64% | 96.72% | 97.02% |
| $b = 2$ | min. | 78.81% | 74.19% | 70.92% | 71.63% | 76.09% |
|  | avg. | 84.11% | 80.84% | 77.83% | 80.47% | 82.06% |
|  | max. | 88.73% | 86.05% | 82.76% | 86.72% | 87.24% |

systems and MU-MISO systems as compared to the proposed method in [15] named as 'Sohrabi-Yu'.

First, we consider a MIMO system with $N = 64$, $N_{RF} = 16$, $M_{RF} = 8$ and $d = 4$. For this system, it is known that the hybrid precoding can achieve the FD performance in the inifinite resolution phase shifter case. The simulation results are presented in Fig. 10 with $b = \infty$ and Fig. 11 with $b = 3$. It is observed from Fig. 10 that, in the inifinite resolution phase shifter case, both the PDD method and the MAP method can achieve the FD performance while there exists a performance gap between the Sohrabi-Yu method and the FD scheme. Furthermore, we find from Fig. 10 that, in the finite resolution phase shifter case with $b = 3$, the PDD method is slightly better than the Sohrabi-Yu method in the low SNR case and almost has the same performance as the Sohrabi-Yu method.However, the MAP method perform worst in this case.

Second, we consider a 4-user MISO system with $N = 64$ and $N_{RF} = 8$. In this case, there is no analog combiner. Hence, the performance loss due to using the finite resolution phase shifter is less than that in the MU-MIMO and MIMO case. The simulation results are presented in Fig. 12 with $b = \infty$ and Fig. 13 with $b = 3$. It is observed from Fig. 12 that the PDD method provides the same performance as the FD scheme in the infinite resolution phase shifter case. Furthermore, as shown in Fig. 13, the PDD method exhibits performance that is very close to the FD performance in the finite resolution phase shifter case, while significantly outperforming the Sohrabi-Yu method.

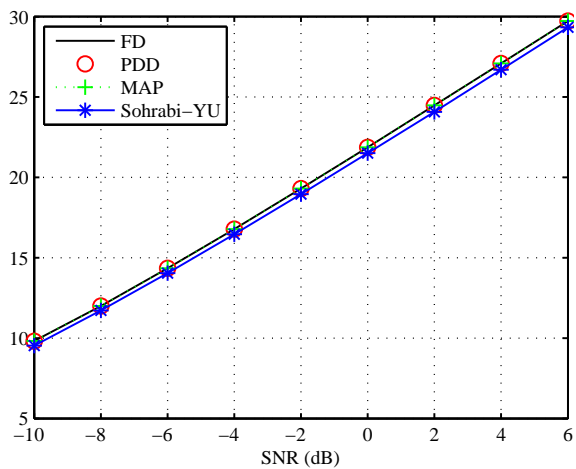

Fig. 10. Spectral efficiency achieved by PDD versus SNR for MIMO system when $N = 64$, $N_{RF} = 16$, $M_{RF} = 8$, $d = 4$, $b = \infty$.

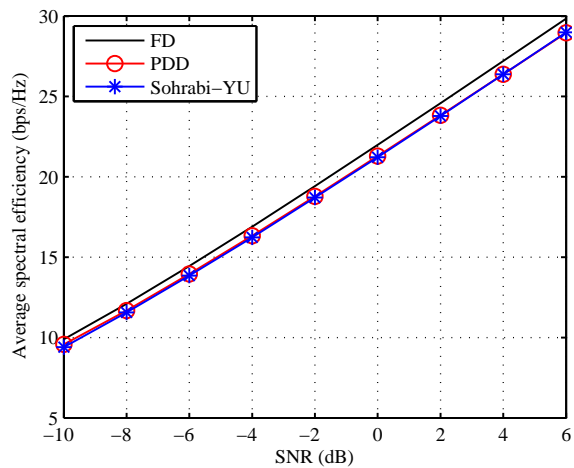

Fig. 11. Spectral efficiency achieved by PDD versus SNR for MIMO system when $N = 64$, $N_{RF} = 16$, $M_{RF} = 8$, $d = 4$, $b = 3$.

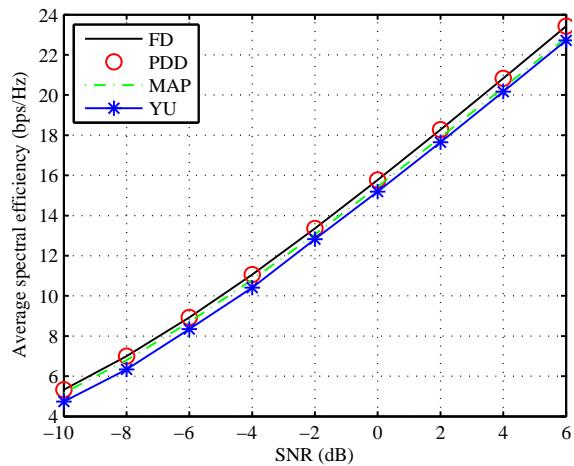

Fig. 12. Spectral efficiency achieved by PDD versus SNR for 4-user MISO system when $N = 64$, $N_{RF} = 8$, and $b = \infty$.

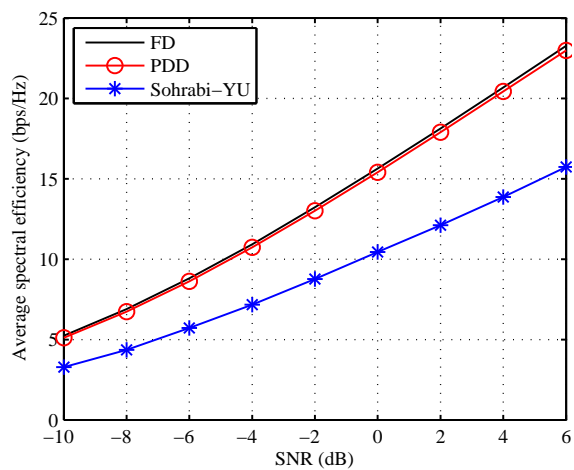

Fig. 13. Spectral efficiency achieved by PDD versus SNR for 4-user MISO system when $N = 64$, $N_{RF} = 8$, and $b = 3$.

## VIII. CONCLUSIONS

By applying penalty dual decomposition method, this paper has proposed an iterative algorithm to address the difficult hybrid precoding problem for mmWave multi-user MIMO systems. Different from the existing hybrid precoding algorithms which are all heuristic, the proposed algorithm has guaranteed convergence to KKT solutions of the hybrid precoding problem. Although the optimality of the proposed hybrid precoding method is not proven, simulation results verify that the proposed hybrid precoding method performs very well and its achievable spectral efficiency is very close to the performance of the fully-digital precoding, implying the capability of hybrid precoding in achieving near-optimal spectral efficiency while greatly reducing the number of RF chains. It is worth mentioning that, all the results are obtained

TABLE IV
ALGORITHM 3: BSUM ALGORITHM

- Input: $\boldsymbol{x}^{k-1}$
- output: $\boldsymbol{x}^k$
0. initialize $\boldsymbol{y}^0 = \boldsymbol{x}^{k-1}$ and set $j = 0$
1. repeat
2.     $\boldsymbol{w} = \boldsymbol{y}^j$
3.     for each $i \in \{1, \ldots, n\}$
4.         $\mathcal{A}_i^j = \arg\min_{\boldsymbol{y}_i \in \mathcal{X}_i} \tilde{\mathcal{L}}_k(\boldsymbol{y}_i; \boldsymbol{w})$
5.         set $\boldsymbol{w}_i$ to be an arbitrary element in $\mathcal{A}_i^j$
6.     end
7.     $\boldsymbol{y}^{j+1} = \boldsymbol{w}$
8.     $j = j + 1$
9. until some termination criterion is met with accuracy $\epsilon_k$
10. $\boldsymbol{x}^k = \boldsymbol{y}^j$

under the assumption of perfect channel state information. In the future, we will consider hybrid precoding with imperfect channel state information.

## APPENDIX A
## A BASIC IMPLEMENTATION OF THE BSUM ORACLE

We outline the basic BSUM algorithm [29] in TABLE IV which can be used to implement the oracle $\boldsymbol{x}^k = \text{BSUM}(P_{\varrho_k, \boldsymbol{\lambda}_k}, \tilde{\mathcal{L}}_k, \boldsymbol{x}^{k-1}, \epsilon_k)$ in Step 2 of the PDD method. Here, $\tilde{\mathcal{L}}_k(\boldsymbol{y}_i; \boldsymbol{w})$ denotes a locally tight upper bound of $\mathcal{L}_k(\boldsymbol{x})$ w.r.t $\boldsymbol{x}_i$ (i.e., the $i$-th block of $\boldsymbol{x}$) at the point $\boldsymbol{w}$.

In the BSUM algorithm [29] applied to a minimization problem with multiple block variables, each time one block variable is cyclicly picked to be optimized while fixing the others by minimizing a locally tight upper bound of the objective. In particular, when the upper bound is simply chosen as the objective function itself, the BSUM algorithm reduces to the BCD method [32]. Hence, the former includes the latter as a special case. On the other hand, note that the BSUM algorithm [29] was proposed for convex constraint cases. When the problem has separable nonconvex constraints, we need to resort to a randomized BSUM algorithm proposed in [28] instead of BSUM for theoretical convergence guarantee. For convenience, we sill refer to it as BSUM algorithm, but keep in mind that the BCD or BSUM algorithm used throughout this paper refers to randomized BSUM algorithm in nonconvex constraint cases, where each time one block variable is randomly picked to be optimized while fixing the others.

## APPENDIX B
## QUADRATIC OPTIMIZATION WITH UNIT MODULUS CONSTRAINTS

This appendix provides an iterative algorithm to address the following quadratic optimization problem with unit modulus constraints:

$$\min_{\mathbf{X} \in \mathcal{M}} \phi(\mathbf{X}) \triangleq \text{Tr}(\mathbf{X}^H \mathbf{A} \mathbf{X} \mathbf{C}) - 2\Re e\left\{\text{Tr}(\mathbf{X}^H \mathbf{B})\right\} \quad (31)$$

where the matrices $\mathbf{A} \in \mathbb{C}^{m \times m}$ and $\mathbf{C} \in \mathbb{C}^{n \times n}$ are positive semidefinite and $\mathcal{M}$ denotes the set of unit modulus constraints, i.e., $|\mathbf{X}(i,j)| = 1, \forall i, j$.

We use BCD-type algorithm to address problem (31) with guaranteed convergence to stationary solutions [32], i.e., in each step we update one entry of $\mathbf{X}$ while fixing the others. Without loss of generality, let us consider the problem of minimizing $\phi(\mathbf{X})$ with respect to $\mathbf{X}(i,j)$ subject to the unit modulus constraint $|\mathbf{X}(i,j)| = 1$, i.e.,

$$\min_{|\mathbf{X}(i,j)|=1} \phi(\mathbf{X}) \quad (32)$$

It is easily known that, the function $\phi(\mathbf{X})$ with respect to $\mathbf{X}(i,j)$ can be expressed as a quadratic function of $\mathbf{X}(i,j)$ in the form of $\tilde{\phi}(\mathbf{X}(i,j)) \triangleq a|\mathbf{X}(i,j)|^2 - 2\Re e\left\{b^*\mathbf{X}(i,j)\right\}$ for some real number $a$ and some complex number $b$. Considering that $|\mathbf{X}(i,j)| = 1$, problem (32) reduces to

$$\max_{|\mathbf{X}(i,j)|=1} \Re e\left\{b^*\mathbf{X}(i,j)\right\}. \quad (33)$$

It follows that the optimal $\mathbf{X}(i,j)$ is equal to $b/|b|$. Hence, in order to update $\mathbf{X}(i,j)$, we only need to know the value of $b$.

In what follows, we show how the complex number $b$ can be easily obtained. First, we have [40]

$$\left.\frac{\partial \tilde{\phi}(\mathbf{X}(i,j))}{\partial \mathbf{X}^*(i,j)}\right|_{\mathbf{X}(i,j) = \tilde{\mathbf{X}}(i,j)} = \frac{1}{2}(a\tilde{\mathbf{X}}(i,j) - b).$$

On the other hand, we have [40]

$$\left.\frac{\partial \phi(\mathbf{X})}{\partial \mathbf{X}^*}\right|_{\mathbf{X} = \tilde{\mathbf{X}}} = \frac{1}{2}(\mathbf{A}\tilde{\mathbf{X}}\mathbf{C} - \mathbf{B}).$$

Combining the above equations, we obtain $[\mathbf{A}\tilde{\mathbf{X}}\mathbf{C} - \mathbf{B}]_{ij} = a\tilde{\mathbf{X}}(i,j) - b$. By expanding $[\mathbf{A}\tilde{\mathbf{X}}\mathbf{C}]_{ij}$ and checking the coefficient of $\tilde{\mathbf{X}}(i,j)$, we have $a\tilde{\mathbf{X}}(i,j) = \mathbf{A}(i,i)\tilde{\mathbf{X}}(i,j)\mathbf{C}(j,j)$. It follows that

$$b = \mathbf{A}(i,i)\tilde{\mathbf{X}}(i,j)\mathbf{C}(j,j) - [\mathbf{A}\tilde{\mathbf{X}}\mathbf{C}]_{ij} + \mathbf{B}(i,j).$$

According to the above analysis, the entries of $\mathbf{X}$ can be recursively updated. The corresponding algorithm for updating $\mathbf{X}$ is summarized in Table III, where the recursion step 5 is due to the fact that $\mathbf{Q}$ should be updated accordingly once $\mathbf{X}(i,j)$ is updated (which is done in step 6). It is easily known that step 5 is the most costly step requiring complexity $O(mn)$. Hence, it can be shown that Algorithm 4 has complexity of $O(I_3 m^2 n^2)$ where $I_3$ denotes the total number of iterations required by Algorithm 4.

## REFERENCES

[1] T. S. Rappaport, S. Sun, R. Mayzus, H. Zhao, Y. Azar, K. Wang, G. N. Wong, J. K. Schulz, M. Samimi, and F. Gutierrez, "Millimeter wave mobile communications for 5G cellular: It will work!" *IEEE Access*, vol. 1, pp. 335–349, 2013.

[2] R. W. Heath, N. Gonzlez-Prelcic, S. Rangan, W. Roh, and A. M. Sayeed, "An overview of signal processing techniques for millimeter wave MIMO systems," *IEEE Journal of Selected Topics in Signal Processing*, vol. 10, no. 3, pp. 436–453, April 2016.

[3] Z. Pi and F. Khan, "An introduction to millimeter-wave mobile broadband systems," *IEEE Communications Magazine*, vol. 49, no. 6, pp. 101–107, June 2011.

TABLE V
ALGORITHM 4: BCD-TYPE ALGORITHM FOR PROBLEM (31)

```
0. set k = 1 and Q = AXC
1. repeat
2.    for (i, j) ∈ {1, 2, ..., m} × {1, 2, ..., n}
3.        b = A(i,i)X(i,j)C(j,j) − Q(i,j) + B(i,j)
4.        x = b/|b|
5.        Q = Q + (x − X(i,j))A(:,i)C(j,:)
6.        X(i,j) = x
7.    end
8.    k = k + 1
9. until some termination criterion is met
```